\documentclass[fleqn,usenatbib]{mnras}

\usepackage{newtxtext,newtxmath}
\usepackage{comment}

\usepackage[T1]{fontenc}

\DeclareRobustCommand{\VAN}[3]{#2}
\let\VANthebibliography\thebibliography
\def\thebibliography{\DeclareRobustCommand{\VAN}[3]{##3}\VANthebibliography}

\usepackage{graphicx}	
\usepackage{amsmath}	
\usepackage[dvipsnames]{xcolor}
\usepackage{multirow}
\usepackage{booktabs}
\usepackage{lscape}

\title[Upper limits on extragalactic civilizations]{Upper limits on transmitter rate of extragalactic civilizations placed by Breakthrough Listen observations}

\author[Yuri Uno et al.]{
Yuri Uno,$^{1}$\thanks{E-mail: yuri.uno@smail.nchu.edu.tw}
Tetsuya Hashimoto,$^{1}$
Tomotsugu Goto,$^{2}$
Simon C.-C. Ho,$^{3}$
Tzu-Yin Hsu, $^{4}$
and Ross Burns$^{5}$
\\
$^{1}$Department of Physics, National Chung Hsing University, No. 145, Xingda Rd., South Dist., Taichung, 40227, Taiwan\\
$^{2}$Institute of Astronomy, National Tsing Hua University, No. 101, Section 2, Kuang-Fu Road, Hsinchu City 30013, Taiwan\\
$^{3}$Research School of Astronomy and Astrophysics, The Australian National University, Canberra, ACT 2611, Australia\\
$^{4}$Department of Physics, National Tsing Hua University, 101, Section 2. Kuang-Fu Road, Hsinchu, 30013, Taiwan (R.O.C.)\\
$^{5}$Department of Science, National Astronomical Observatory of Japan, 2-21-1 Osawa, Mitaka, Tokyo 181-8588, Japan
}

\date{Accepted 2023 March 28. Received 2023 March 20; in original form 2022 December 7}

\pubyear{2023}

\begin{document}
\label{firstpage}
\pagerange{\pageref{firstpage}--\pageref{lastpage}}
\maketitle

\begin{abstract}
The Search for Extra-Terrestrial Intelligence (SETI) has been conducted for over sixty years, yet no technosignatures have been identified. Previous studies have focused on stars in our galaxy, with few searches in the extragalactic Universe despite a larger volume being available. Civilizations capable of harvesting energy from a star or a galaxy are classified as KII or KIII on the Kardashev scale, respectively. Technosignatures from such advanced civilizations would be extremely luminous and detectable by current radio telescopes, even from distant galaxies. To explore the frontier of extragalactic SETI, we investigate the likely prevalence of extragalactic civilizations possessing a radio transmitter, known as the transmitter rate, based on observational results from the Breakthrough Listen (BL) observations. We calculated the transmitter rate by considering the background galaxies in the field of view of target stars in BL observations. We used a statistical method to derive the total mass of stars in those background galaxies from a galaxy stellar mass function. 
Our statistical method suggests that less than one in hundreds of trillions of extragalactic civilizations within 969 Mpc possess a radio transmitter above 7.7$\times$10$^{26}$ W of power, assuming one civilization per one-solar-mass stellar system. Additionally, we cross-matched the BL survey fields with the WISE$\times$SuperCOSMOS Photometric Redshift Catalogue and compared with the statistical method.  Our result sets the strictest limits to date on the transmitter rate at such high power levels, emphasizing the high efficiency of searching for radio transmitters in galaxies and
the rarity of technologically advanced civilizations in our Universe.

\end{abstract}




\section{Introduction}
The Search for Extra-Terrestrial Intelligence (SETI) is a growing field in astronomy which aims to find evidence of extraterrestrial civilizations (ETC) that possess communicative technology in the Universe \citep{2022AcAau.199..166H, 2022AcAau.190...24W}. SETI has been conducted for more than six decades since \citet{1959Natur.184..844C} proposed that radio transmission from ETC might be detectable. The first search of such radio transmission was conducted by \citet{1961PhT....14d..40D} where he observed two stars located ten light years away. Assuming that ETCs will reside in a stellar system, most of the previous searches have focused on the nearby Universe within our Galaxy. So far, technological signatures from ETCs, also called technosignatures, have not been identified. No detection does not mean that we are alone in the Universe. \citet{2018AJ....156..260W} claim that the search efforts have not been enough, comparing the ratio of the searching volume with that of a small swimming pool of water and the sea. 

\citet{2017AJ....153..110G} pioneered searching for radio transmissions from ETCs in the extragalactic universe, in which they can efficiently look for signals better than targeting individual stars in our galaxy as one pointing in a radio observation may include billions of stars. 
Their search towards nearby galaxies, M31 and M33, found no radio transmission at the frequency of the neutral hydrogen line, which is one of the frequencies known as a ``magic frequency''.
This frequency could have a higher chance to be detected by other civilizations \citep{1993ASPC...47..161G}. Nevertheless, the result has set one of the strictest limits on the prevalence of radio transmitters, known as the transmitter rate, above 2.1$\times$10$^{21}$ W.

Another way to efficiently search for radio transmissions is by considering all stars within a field of view (FoV) of a radio telescope, i.e., not only the target star at the centre. The FoV is the solid angle where a single-dish radio telescope has sensitivity. The angle of FoV is much larger than the apparent size of the target star. Therefore, one target observation simultaneously observes many other astronomical sources in the background of the target stars. \citet{2020MNRAS.498.5720W} investigated such background stars in Breakthrough Listen (BL) observations led by \citet{2017ApJ...849..104E} and \citet{2020AJ....159...86P}, cross-matching the FoVs of target stars with the {\it Gaia} DR2 catalogue. As a result, \citet{2020MNRAS.498.5720W} found about three hundred thousand background stars within 10 kpc, which is two hundred times the number of stars originally targeted by BL observations.
 
The Kardashev scale is a measure of the level of advancement of ETC based on their power consumption \citep{1964SvA.....8..217K}. KI, KII, and KIII civilizations have power consumption equivalent to a planet, a star, and a galaxy, respectively. If a civilization is as advanced as KII or KIII, its radio transmission would be able to propagate cosmological distances. The pioneering work of searching for extragalactic transmitters was conducted by 
\citet{2023MNRAS.519.4581G}, 
and they focused on extragalactic sources within the FoV of BL observations led by \citet{2017ApJ...849..104E} and found more than a hundred thousand extragalactic sources using the NASA/IPAC Extragalactic Database (NED), including rather exotic sources ranging from active galactic nuclei (AGN) to a gravitational lensing system. \citet{2023MNRAS.519.4581G} assumed $\sim10^{11}$ M$_{\odot}$ for each extragalactic source in the FoV to derive the upper limits on the transmitter rate. However, this could significantly overestimate the number of stars because $10^{11}$ M$_{\odot}$ is comparable to the highest galaxy stellar mass in the local Universe.

This paper investigates the upper limits on the transmitter rate of extragalactic civilizations using better estimates on galaxy stellar mass, i.e., the number of stars, by employing a galaxy stellar mass function and empirically derived galaxy stellar mass for individual galaxies in the FoV of the BL observations.

Section \ref{sec:data} summarises the BL project and data, while section \ref{sec:method} explains the methods of our analysis. In section \ref{sec:result}, we report the results of our analysis using a stellar mass function. Section \ref{sec:discussion} discusses the results, and Section \ref{sec:conclusion} presents our conclusion.




\section{Breakthrough Listen Project}
\label{sec:data}
The Breakthrough Listen (BL) project is conducting the most comprehensive search for technosignature using the world's largest radio telescopes, such as the Robert C. Byrd Green Bank Telescope (GBT) and the Parkes radio telescope (\citealt{2017PASP..129e4501I}). BL is searching for a drifting narrow-band radio signal with a signal width of a few Hz frequencies. Although it is uncertain if the radio transmission would be a narrow band, if such a signal was detected, we could distinguish it from natural radio sources as the most narrow band natural sources, such as masers, typically have widths of about a few kHz. Doppler drifting in frequency against time, caused by the relative motion between the transmitter and receiver, is another way of distinguishing technosignatures from natural radio sources. 

Among the BL observation campaigns, \citet{2020AJ....159...86P} searched for a narrowband radio transmission towards the largest number of targets. \citet{2020AJ....159...86P} observed 1327 nearby stellar systems at L- and S-bands using GBT and Parkes (hereafter, we denote GBT observation at L- and S-bands as GBT-L, GBT-S, and Parkes observation at S-band as PAR-S). The observations were conducted from January 2016 to March 2019, and the raw data was processed into spectral data with a frequency and time resolution of approximately 2.79 Hz and 18.25 s,
respectively. The generated data were analyzed with the {\it turbo}SETI software with drift rates between $\pm$4 Hz s$^{-1}$ and a detection threshold above signal-to-noise ratio (SNR) of $>10$. {\it turbo}SETI detected twenty thousand technosignature candidates in the data, which were later rejected as radio frequency interference (RFI). Since a large number of targets were observed, the FoVs of BL targets observed by \citet{2020AJ....159...86P} provide the best opportunity for us to constrain the upper limit of the prevalence of extragalactic transmitters.

The FoVs of GBT-L, GBT-S, and PAR-S according to \citet{2020MNRAS.498.5720W}, were 8.4$\arcmin$, 5.5$\arcmin$, and 6.4$\arcmin$, respectively. An FoV is equivalent to the diameter of the surface area projected onto the sky, and for the purposes of this study, we define each surface area observed by BL as
a BL field, following the terminology of \citet{2023MNRAS.519.4581G}. The total number of unique BL fields observed by \citet{2020AJ....159...86P} were 880, 998, and 185 fields for GBT-L, GBT-S, and PAR-S, respectively. To exclude the noisy environment around the galactic plane, we only consider BL fields at $|b|\leq 20^\circ$, resulting in 636, 694, and 109 BL fields for GBT-L, GBT-S, and PAR-S, respectively, for further analysis. 

\section{Method}
\label{sec:method}
\subsection{Detectability of extragalactic transmitters}
The minimum detectable specific flux of a radio telescope is defined as,
\begin{equation}
    F_{\rm min} = SNR_{\rm min} \times SEFD \times \sqrt{\frac{BW}{n_{\rm pol}\times t_{\rm obs}}}\times \Delta f_t,
	\label{eq:sensitivity}
\end{equation}
where $F_{\rm min}$ is in units of W m$^{-2}$, assuming a transmitting signal $\Delta f_t$ 
has 1~Hz width following \citet{2020AJ....159...86P}, $SNR_{\rm min}$ is the minimum signal-to-noise ratio, $SEFD$ is system-equivalent flux density in Jy, $BW$ is the frequency bandwidth in Hz, $n_{\rm pol}$ is the number of polarizations, and $t_{\rm obs}$ is the integration time of observation in seconds.

The least detectable power of a radio transmitter, minimum Equivalent Isotropic Radiated Power (${\rm EIRP}_{\rm min}$)  in watts, at a distance $d_{\rm lim}$ is given as
\begin{equation}
    {\rm EIRP}_{\rm min} = 4\pi d_{\rm lim}^2 F_{\rm min},
      \label{eq:transmitter-power}
\end{equation}
where $d_{\rm lim}$ is the luminosity distance between the radio transmitter and the receiver.  

The transmitters of extragalactic civilizations would need great power for their signals to be detected by others across intergalactic distances. To handle such great power, civilizations should be very advanced. The advanced degree relates to the civilizations power consumption level, and is called the Kardashev scale \citep{1964SvA.....8..217K}. A simplified Kardashev scale $K$ is defined as
\begin{equation}
    K = \frac{\log{P}-6}{10},
      \label{eq:kardashev-scale}
\end{equation}
where $P$ is the power of the radio transmitter in watts \citep{sagan2000carl}. If a radio telescope has $F_{\rm min}$ of around 20~Jy, it could detect the radio signal from KII-type civilizations ($P\simeq2.4\times10^{27}$~W), at distances of up to 1000~Mpc. 

\subsection{Number of stars in the BL fields}
\label{sub:numberofstars}
We describe two methods to estimate the total stellar mass in the FoV of BL observations in this section. The total stellar mass is then converted to the total number of stars used in Section \ref{Transmitter_Rate}, assuming 1 M$\odot$ for a star.

\subsubsection{Statistical method using a galaxy stellar mass function}
\label{method1}
The number of stars in the BL fields can be calculated by multiplying the search volume of BL observations with the volume density of stars. The total search volume of BL observations is given as,
\begin{equation}
   V_r = f_s \frac{4\pi d_{\rm lim}^3}{3} N_{\rm target},
    \label{eq:volume}
\end{equation}
\noindent
where $V_r$ is the total search volume,  which differs for each receiver as the size of FoVs is different, $f_s$ is the fractional surface area of a BL field, and $N_{\rm target}$ is the number of targets. In this work, we adopt $d_{\rm lim}=$ 969, 980, and 930 Mpc for GBT-L, GBT-S, and PAR-S, respectively.
These values are calculated by median values of distances to wiseSCOS galaxies within the BL fields, as described in Section \ref{method2}. 

The volume density of stars can be statistically calculated from the galaxy stellar mass function. Here we use the galaxy stellar mass function defined by \citet{2012MNRAS.421..621B} to derive the density $\rho$ as follows,
\noindent
\begin{equation}
\rho=\int^{M_{\rm max}}_{M_{\rm min}} M \exp{\bigg(\frac{-M}{M^*}}\bigg) \bigg( \phi_1^* \big(\frac{-M}{M^*}\big)^{\alpha_1} + \phi_2^* \big( \frac{M}{M^*}\big)^{\alpha_2} \bigg) \frac{dM}{M^*},
\label{eq:stellar-density} 
\end{equation}
\noindent
where $M$ is the stellar mass of galaxies, and the parameter values cited from \citet{2022MNRAS.513..439D}, 
$\log{M^*}=10.745\pm0.020$, $\log{\phi_1^*}=-2.437\pm0.016$, $\log{\phi_2^*}=-3.201\pm0.064$, $\alpha_1=-0.466\pm0.069$, $\alpha_2=-1.530\pm0.027$, $M_{\rm min}=10^6M_\odot$, and $M_{\rm max}=10^{12}M_\odot$. 
The derived stellar mass density in logarithmic scale is $\log{\rho}=8.473\pm0.003$ ${M_\odot}$ Mpc$^{-3}$. 
This statistical error was estimated by considering the uncertainties of the best-fit parameters of the galaxy stellar mass function as reported by \citet{2022MNRAS.513..439D}. We conducted Monte Carlo simulations by randomly sampling the best-fit parameters assuming Gaussian probability density functions with standard deviations equal to the parameter errors. The Monte Carlo simulations were repeated 10$^{4}$ times to derive the standard deviation of the stellar mass.

Because the statistical error is negligibly small, i.e., 0.003 dex, we hereafter include systematic errors due to different assumptions on the initial mass function \citep[IMF;][]{2014ARA&A..52..415M}. 
The derived total stellar masses for GBT-L, GBT-S, and PAR-S are $\log (M_{\rm gal} /M_{\odot})$ = $14.46_{(-0.03)}^{(+0.18)}$, $14.15_{(-0.03)}^{(+0.18)}$, and $13.41_{(-0.03)}^{(+0.18)}$, respectively. Here, these uncertainty values in the parentheses are systematically derived from the different IMFs, \citet{2003PASP..115..763C}, \citet{2001MNRAS.322..231K}, and \citet{1955ApJ...121..161S} for the lower bound, nominal value, and upper bound, respectively.

These numbers are then converted from the total stellar mass to the total number of stars assuming 1 $M_{\odot}$ for a star, which gives the same number as the number of stars in the BL fields.

\subsubsection{Crossmatching method with an extragalactic source catalogue}
\label{method2}
The search volume of BL is not so large when $d_{\rm lim}$ is less than one thousand~Mpc, making it possible to conduct a crossmatch of the BL fields with an extragalactic source catalogue. We use the WISExSuperCOSMOS Photometric Redshift Catalogue \citep[wiseSCOS,][]{2016ApJS..225....5B} to investigate the extragalactic sources in BL fields. 

wiseSCOS is a galaxy catalogue with twenty million galaxies at redshifts up to 0.6 and a median redshift of $z=0.2$ \citep{2016ApJS..225....5B}. 
The distance to each galaxy in the wiseSCOS catalogue is estimated from the photometric redshift, and its uncertainty ($\sigma_z$) is given as,
\begin{equation}
    \sigma_z = 0.033 (1+z).
    \label{uncertainty_redshift}
\end{equation}
The photometric redshift has a large uncertainty compared with the spectroscopic redshift. Therefore, the number of galaxies around the boundary of a certain survey depth can fluctuate due to the redshift error. 
To take this effect into account, we applied a Monte Carlo simulation and recalculated the distance considering the uncertainty, and then identified the galaxies in BL fields within $d_{\rm lim}$~Mpc. The mass of each crossmatched galaxy is calculated from the empirical relationship between stellar mass and WISE photometry, which was proposed by \citet{2014ApJ...782...90C}. Following their calculation, the equation is given as,
\noindent
\begin{equation}
    M_{\rm star} = 10^{-0.4 (M_{\rm abs}-M_{\rm sun})} 10 ^{- 1.96 (W_{\rm 3.4\mu m} - W_{\rm 4.6 \mu m}) - 0.03},
\end{equation}
\noindent
where $M_{\rm sun}=3.24$ and, $M_{\rm abs}$ is the absolute magnitude of the galaxy defined as,
\begin{equation}
    M_{\rm abs} = M_{\rm app} - 5 \log_{10}{(d_{\rm pc})} + 5.
\end{equation}
$M_{\rm app}$ is the apparent magnitude of the galaxy in the $W_{3.4{\rm \mu m }}$ band and $d_{\rm pc}$ is the distance to galaxy in pc. 
We iterated these processes over 10$^4$ times, and each process provided the total stellar mass within the BL fields.
Histograms of the total stellar mass were fitted with a Gaussian function to estimate the statistical uncertainty due to the photometric-redshift error.
The derived total stellar masses for GBT-L, GBT-S, and PAR-S were $\log (M_{\rm gal} /M_{\odot}) = 13.63\pm0.01^{(+0.18)}_{(-0.03)}, 13.11\pm0.02^{(+0.18)}_{(-0.03)}$, and $12.53\pm0.05^{(+0.18)}_{(-0.03)}$, respectively.
Here, the systematic errors due to different assumptions on IMFs are shown in parenthesis (see Section \ref{method1} for details).
These numbers are identical to the numbers of stars in the BL fields, assuming 1 $M_{\odot}$ for a star.

\subsection{Transmitter Rate}
\label{Transmitter_Rate}
The transmitter rate is a metric to evaluate the comprehensiveness of a SETI search. The upper limit on the transmitter rate is defined by \citet{2017ApJ...849..104E} as,
\noindent
\begin{equation}
TR = \frac{1}{N_{\rm star}\nu_{\rm rel}},
    \label{eq:trasmitter_rate}
\end{equation}
\noindent
where $\nu_{\rm rel}$ is the relative frequency defined as the total bandwidth divided by the observation frequency. The transmitter rate is, in other words, the fractional upper limit of stellar systems possessing a radio transmitter in a certain frequency parameter space.

\section{result}
\label{sec:result}
\subsection{Statistical method}
\label{subsec:statistical method}
The GBT-L, GBT-S, and PAR-S have total surveyed surface areas of 9.79, 4.58, and 0.97 deg$^2$, respectively. The surveyed volumes $V_{\rm r}$ are determined by the assumed distance to the transmitter, represented by the equation $V_{\rm r}= k {d_{\rm lim}}^3$, where $k$ is a dimensionless constant. The $k$ values for GBT-L, GBT-S, and PAR-S are $1.0\times 10^{-3}$, $4.7\times10^{-4}$, and $9.9\times10^{-5}$, respectively. For example, when the distance is 100 Mpc, the search volume of GBT-L is approximately $10^3$ Mpc$^3$. The corresponding stellar mass and transmitter rates are summarized in Tab. \ref{tab:result}.


\subsection{Crossmatching method}
\label{subsec:crossmatching method}
In the BL fields, there are a total of 2605, 717, and 211 galaxies crossmatched with wiseSOCS for GBT-L, GBT-S, and PAR-S, respectively, regardless of their distance. Tab. \ref{tab:result} summarizes the stellar mass and transmitter rate results when the matched galaxies' median distance is used as $d_{\rm lim}$. We obtained a significantly lower number of matched galaxies compared to \citet{2023MNRAS.519.4581G} because we only used wiseSOCS galaxies with known redshift information, while \citet{2023MNRAS.519.4581G} included many extragalactic sources with unknown redshifts.


\begin{table*}
	\centering
	\caption{The transmitter rates obtained from different BL surveys are presented in this table. The first row indicates the depth of the survey. The second row shows the number of galaxies within the survey volume. The third row displays the relative frequency $\nu_{\rm rel}$, where $f$ represents the observed frequency , and $BW$ denotes the total bandwidth in GHz. The fourth row shows the total stellar mass $M_{\rm gal}$ of BL fields in terms of $d_{\rm lim}$ on a common logarithmic scale. The fifth row is the minimum Equivalent Isotropic Radiated Power (EIRP$_{\rm min}$) defined by Eq. \ref{eq:transmitter-power}. 
    The sixth row is the transmitter rate in the logarithmic scale derived by Eq. \ref{eq:trasmitter_rate}. 
    Each uncertainty value represents a standard deviation derived by Monte Carlo simulations described in Section \ref{method2} except for errors in parenthesis which indicate systematic errors due to different assumptions on initial mass functions \citep[IMF;][]{2014ARA&A..52..415M}. 
    The upper (lower) bound of the systematic error corresponds to the IMF of \citet{1955ApJ...121..161S} \citep{2003PASP..115..763C}, whereas the nominal value of the stellar mass assumes the IMF by \citet{2001MNRAS.322..231K}.
    Note that the statistical errors of the stellar mass and the transmitter rate for the statistical method are negligibly small compared to the systematic errors (see Section \ref{method1} for the details).
    }
 
	\label{tab:result}
    \begin{tabular}{rrrrrrr}
\hline
\multicolumn{1}{r|}{}              & \multicolumn{2}{c|}{GBT-L} & \multicolumn{2}{c|}{GBT-S} & \multicolumn{2}{c|}{PAR-S} \\ \hline \hline
                    & statistical & crossmatch & statistical & crossmatch & statistical & crossmatch \\
$d_{\rm lim}$ (Mpc) & 969         & 969        & 980         & 980        & 930         & 930        \\ \\
$N_{\rm gal}$       & $(1.9\pm0.18)\times10^{5}$  & 1305$\pm$15   & $(9.1\pm0.16)\times10^{4}$   & 355$\pm$8     & $(1.7\pm0.88)\times10^{4}$   & 106$\pm$4  \\ \\
$\nu_{\rm rel} (BW/f )$       & 0.8/1.5     &  0.8/1.5   & 1.0/2.35      &  1.0/2.35     & 0.85/3.025       & 0.85/3.025   \\ 
\\
$\log{M_{\rm gal}}$ ($M_{\odot}$) & $14.46^{(+0.18)}_{(-0.03)}$   & $13.63\pm$0.01$^{(+0.18)}_{(-0.03)}$    & $14.15^{(+0.18)}_{(-0.03)}$  & 13.11$\pm$0.02$^{(+0.18)}_{(-0.03)}$   & $13.41^{(+0.18)}_{(-0.03)}$    & 12.53$\pm$0.05$^{(+0.18)}_{(-0.03)}$   \\
\\
${\rm EIRP_{\rm min}\, (WHz})$       & 7.72$\times 10^{26}$    & 7.72$\times 10^{26}$   & 7.89  $\times 10^{26}$    & 7.89  $\times 10^{26}$     & 2.42  $\times 10^{27}$    & 2.42  $\times 10^{27}$ \\ \\
$\log{\rm TR}$       & 
$-14.15^{(+0.03)}_{(-0.18)}$    & $-13.35\pm0.01^{(+0.03)}_{(-0.18)}$ & $-13.77^{(+0.03)}_{(-0.18)}$    & 
$-12.75\pm0.02^{(+0.03)}_{(-0.18)}$ 
& $-12.85^{(+0.03)}_{(-0.18)}$     & 
$-11.98\pm0.05^{(+0.03)}_{(-0.18)}$  

\\ \hline
\end{tabular}
\end{table*}

\begin{figure*}
	\includegraphics[width=\textwidth]{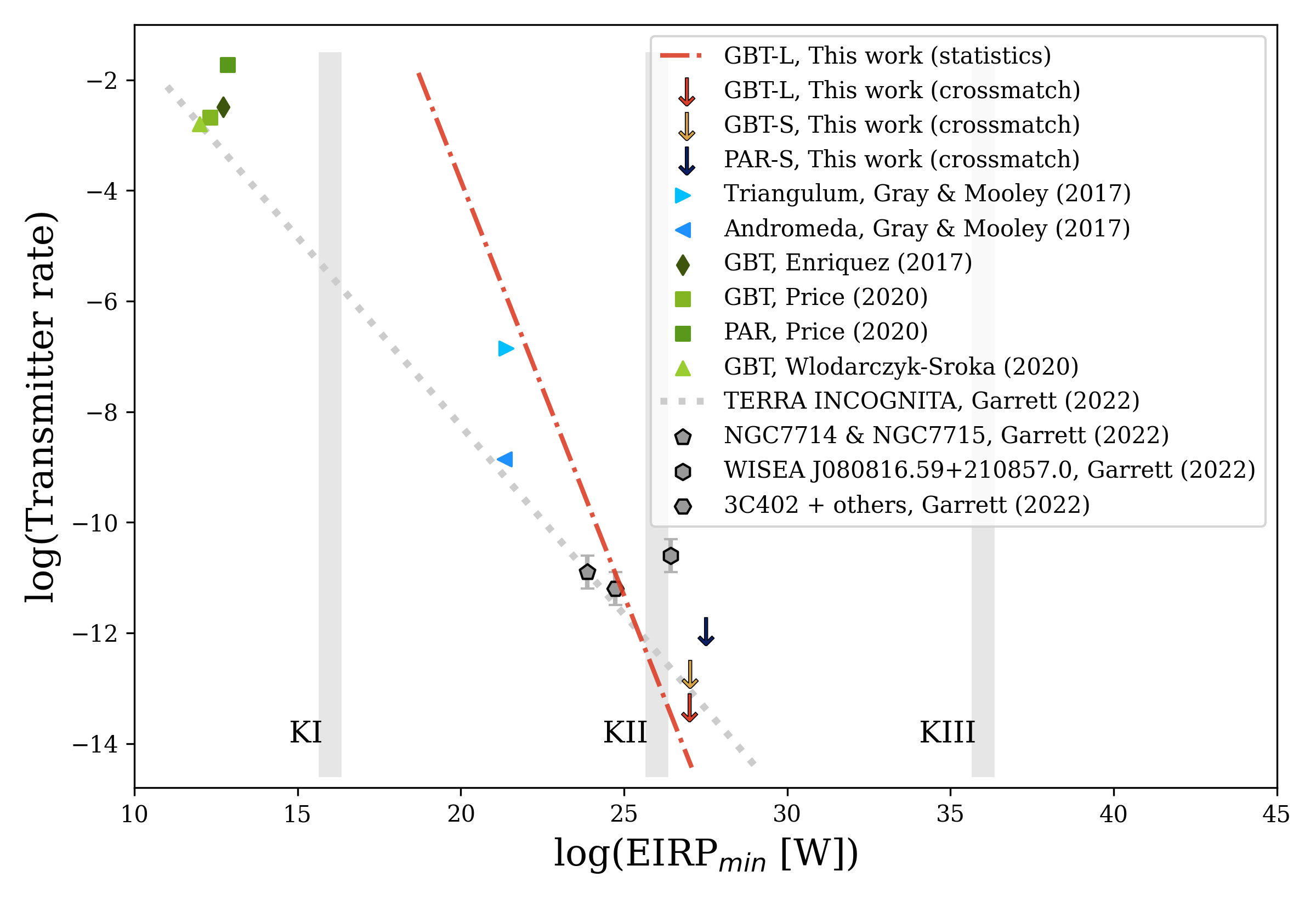}
    \caption{Upper limits on the transmitter rates as a function of EIRP$_{\rm min}$ in this work and  previous studies are presented. The dot-dash lines show our result for the case of GBT-L, using the statistical method described in Section \ref{method1}. 
    The results based on the crossmatching method are shown in arrows, with different colours corresponding to different
    surveys.
    Other markers in different colours represent the upper limits presented in previous studies \citep{2017AJ....153..110G, 2017ApJ...849..104E, 2020AJ....159...86P, 2020MNRAS.498.5720W, 2023MNRAS.519.4581G}.
    The grey dotted line connecting the previous most constraining values reported by 
    \citet{2020MNRAS.498.5720W} and \citet{2023MNRAS.519.4581G} was introduced as terra incognita \citep{2023MNRAS.519.4581G}. 
    The terra incognita indicates the line below which the parameter space needs to be explored at better sensitivity and a broader search area.}
    \label{fig:transmitter-rate}
\end{figure*}

\section{discussion}
\label{sec:discussion}
Previous extragalactic SETI research has focused solely on particular galaxies or galactic sources \citep{2017AJ....153..110G, 2023MNRAS.519.4581G}. For the first time, we have taken into account the total stellar mass of galaxies within the FoV of BL targets and set an upper limit on the existence of extragalactic transmitters. Fig. \ref{fig:transmitter-rate} compares our transmitter rates with those of previous works. Our analysis has statistically accounted for hundreds of thousands of background galaxies up to $\sim$900--1000 Mpc in BL fields, which contains a stellar mass equivalent to several orders of magnitude more Sun-like stars than previous works. 
Therefore, the upper limits on the transmitter rate derived in this work are $\sim$ 1--2 orders of magnitude more stringent than those in the previous work of \citet{2023MNRAS.519.4581G}. 
 
The results of our statistical and crossmatching method are reasonably consistent. The statistically derived upper limits are stricter, 
as we can include stars in the relatively smaller galaxies (10$^{6} M_{\odot}$), which are likely not sampled in the wiseSCOS. 
For GBT-L, which has the highest search volume, we found that the logarithmic scale mass of galaxies within the search volume is 14.46 $M_\odot$ using the statistical method. Assuming that each civilization is associated with one solar system, the probability of a civilization having a powerful radio transmitter or power above 7.7$\times 10^{26}$ W existing is less than one in several hundred trillion.

\citet{2023MNRAS.519.4581G} defined a terra incognita as the limit of our current search for radio transmitters by fitting a linear function to the previous two stringent constraints \citep{2020MNRAS.498.5720W,2023MNRAS.519.4581G}. The intersection of the terra incognita and our statistically derived transmitter rate in Fig. \ref{fig:transmitter-rate} is the turning point at which we benefit more from blind search observations than from targeting individual stars or galaxies. 
Therefore, the attribution of the background galaxies becomes more significant at the turning point.

\section{conclusion}
\label{sec:conclusion}
 If there is an advanced civilization emitting a high-power radio transmitter, such as can be expected for a KII or KIII type civilization, radio telescopes on Earth have the potential to detect such signals 
 even from an extragalactic distance. 
 We investigated the total mass of the galaxies in the background within the FoV of BL targets observed simultaneously during BL observations. 
 Based on the statistical (crossmatch) method, we 
 derived upper limits on the transmitter rate of $\log {\rm TR}=-14.15^{(+0.03)}_{(-0.18)}~\big(-13.35\pm0.01^{(+0.03)}_{(-0.18)}\big), -13.77^{(+0.03)}_{(-0.18)}~\big(-12.75\pm0.02^{(+0.03)}_{(-0.18)}\big)$, and $-12.85^{(+0.03)}_{(-0.18)}~\big(-11.98\pm0.05^{(+0.03)}_{(-0.18)}\big)$ in GBT-L, GBT-S, and PAR-S fields, respectively.
 These are the strictest upper limits on the transmitter rate at the energy rate of $\sim10^{27}$ W, i.e., KII civilization on the Kardashev scale. 
 Currently, the transmitter rate is one of the few scientific ways of evaluating the prevalence of technological civilizations.
 Our result suggests that the existence of KII-type civilizations are extremely limited, and the search for radio transmissions should continue. Our results also highlight the importance of targeting not only nearby stars in the Milky Way Galaxy but also in other galaxies in the Universe. 

\section*{Acknowledgements}
\label{sec:acknowledgements}
We acknowledge the anonymous referee for their helpful comments in improving this paper. YU thanks the university for supporting research activities through the NCHU scholarship. YU would also like to thank many discussions during the first Penn State SETI conference in 2022, especially Prof. Michael Garrett, for carefully paying attention to this research. YU would also like to extend thanks to Dr. Shotaro Yamasaki for providing valuable suggestions and advice, which significantly improved the content of this paper. Furthermore, YU is grateful to Dr. Danny Price for kindly sharing the BL data and patiently answering many questions regarding BL observations. TH acknowledges the support of the National Science and Technology Council of Taiwan through grants 110-2112-M-005-013-MY3, 110-2112-M-007-034-, and 111-2123-M-001-008-. TG acknowledges the support of the National Science and Technology Council of Taiwan through grants 108-2628-M-007-004-MY3 and 111-2123-M-001-008-. This research made use of Astropy, a community-developed core Python package for Astronomy \citep{Astropy2018}.

\section*{Data Availability}
The data underlying this article are available at \url{https://seti.berkeley.edu/listen2019/listen2019_tables.pdf} for the BL observations and at \url{http://ssa.roe.ac.uk/WISExSCOS.html} for the wiseSCOS data.

\bibliographystyle{mnras}
\bibliography{ref} 

\bsp	
\label{lastpage}
\end{document}